# On the (Mis)Use of Machine Learning with Panel Data

Augusto Cerqua[†], Marco Letta[‡], Gabriele Pinto[§]

First version: November 2024

This version: May 2025

**Abstract**

*We provide the first systematic assessment of data leakage issues in the use of machine learning on panel data. Our organizing framework clarifies why neglecting the cross-sectional and longitudinal structure of these data leads to hard-to-detect data leakage, inflated out-of-sample performance, and an inadvertent overestimation of the real-world usefulness and applicability of machine learning models. We then offer empirical guidelines for practitioners to ensure the correct implementation of supervised machine learning in panel data environments. An empirical application, using data from over 3,000 U.S. counties spanning 2000-2019 and focused on income prediction, illustrates the practical relevance of these points across nearly 500 models for both classification and regression tasks.*

**JEL-Codes:** C33, C53.

**Keywords**: machine learning, prediction policy problems, panel data, data leakage.

**Replication package**: all data and codes for replication are publicly available at this link.

Acknowledgments: Augusto Cerqua and Marco Letta gratefully acknowledge financial support from the Italian PRIN 2022 grant for the project "Strengthening TARgeting and Guidance with Actionable and Timely Evidence (STARGATE) for the implementation of the Italian National Recovery and Resilience Plan", CUP: F53D23003220006.

[†] Department of Social Sciences and Economics, Sapienza University of Rome, Italy. Email: augusto.cerqua@uniroma1.it
[‡] Department of Social Sciences and Economics, Sapienza University of Rome, Italy. Email: marco.letta@uniroma1.it
[§] Department of Social Sciences and Economics, Sapienza University of Rome, Italy. Email: gabriele.pinto@uniroma1.it

## 1. Introduction

In recent years, economics and other social sciences have enthusiastically embraced the use of machine learning (ML) to address "prediction policy problems" (Kleinberg et al., 2015) and improve policy targeting. Since the issue of *whom to target* arises in many settings (Athey et al., 2025), numerous scholars have started to apply supervised ML algorithms to panel data to analyze and support policy targeting and design. Topics covered include enhancing investments in energy efficiency and assessing the costs of energy policy (Christensen et al., 2024; Jarvis et al., 2022), increasing the effectiveness of tax audits (Battaglini et al., 2024), and improving the forecasting and targeting of latent outcomes at both the local (e.g. bankruptcy, corruption) and national levels (e.g. financial crises, asylum seeker flows, individuals in need of income support) (Antulov-Fantulin et al., 2021; Ash et al., 2024; Boss et al., 2024; Bluwstein et al., 2023; de Blasio et al., 2022; Sansone & Zhu, 2023).

The overall idea underlying papers in this new tradition is simple and compelling: to leverage increasingly available rich panel datasets for accurately predicting complex social phenomena on new, previously unseen data, thus offering policymakers predictions that can be used to refine policies and shape outcomes in the desired direction (Kleinberg et al., 2015). In practice, the main purpose of most of these studies is to identify the areas or units most at risk of undesirable outcomes or susceptible to a given hard-to-detect phenomenon. Accordingly, the ML algorithms are asked to find recurrent patterns in the data and predict where the phenomenon under analysis is most likely to occur. By pinpointing hotspots or 'red flags' (e.g. high-risk areas), these models can help policymakers allocate resources in a cost-efficient way, ensuring that policy efforts are concentrated where they are needed most and allowing for proactive rather than reactive measures.

Although we believe that the insights provided by this new and active research area are valuable and that ML has much to offer in improving policy targeting and design, we believe it is important to raise awareness about common mistakes associated with the default application of supervised ML to panel data, such as the use of contemporaneous covariates, i.e., covariates observed at time *t* used as 'forbidden' predictors to forecast outcomes at time *t* (Petropoulos et al., 2022), or the split of the observations into training and testing sets in a way that does not make them truly disjoint (Kapoor & Narayanan, 2023). Such modeling choices can lead to data leakage, defined as the unintended use of information during model training and validation that would not be available at the prediction stage, which has been deemed 'one of the top ten data mining mistakes' (Kaufman et al., 2012).[1] The presence of data leakage in an ML pipeline can lead to overly optimistic measures of out-of-sample performance, which, in turn, can result in misleading policy prescriptions and overconfidence in the ML's actual ability to support policy efforts.

The perils of data leakage associated with the use of ML have recently come under intense scrutiny across many different scientific fields, including biology, medicine, computer security, peace studies, nutrition, and satellite imaging (Apicella et al., 2024; Bernett et al., 2024; Kapoor & Narayanan, 2023; Rosenblatt et al., 2024). Data leakage has even been indicated as one of the main culprits of the reproducibility crisis in ML-based science (Kapoor & Narayanan, 2023), potentially contributing to producing 'illusions of understanding' in AI-driven scientific research (Messeri & Crockett, 2024). In economics, however, these issues are frequently overlooked by practitioners in applied work, despite the fact that empirical analyses increasingly rely on the combination of ML and panel data, which are particularly prone to data leakage, and despite econometricians typically being fully aware of these issues and adjusting their sampling strategies accordingly (e.g., Babii et al., 2023; Babii et al., 2024). A recent examination of the disconnect between econometric theory and empirical practice and its severe consequences in terms of temporal leakage in trading and finance forecasting—unrelated to either the use of ML or panel data—is provided by the replication study of Cakici et al. (2024).

In this work, we aim to bridge this disconnect by conducting the first comprehensive assessment of the causes and consequences of data leakage when applying ML algorithms to panel data. We start by providing an organizing framework that clarifies the conceptual and practical pitfalls associated with the uncritical use of supervised ML with panel data. We then propose empirical guidelines for practitioners to ensure the correct implementation of supervised ML techniques in practically relevant panel data environments, emphasizing the need to clarify the primary goal *ex-ante* and align the ML analysis accordingly. To make our methodological discussion concrete, we illustrate these points with an application based on a balanced panel dataset of over 3,000 U.S. counties focusing on both a classification problem and a regression problem. Our target reader is the applied economist for whom this paper can serve as an entry point to correctly harness these powerful tools on their panel data.

For this analysis, we focus on aggregate panel data, also known as 'time-series cross-sectional data,' as the benchmark case. This type of data involves observations of numerous administrative entities (such as regions or countries) across multiple time periods and is increasingly used in studies on policy targeting. In addition, by analyzing aggregate panel data, we can underline their specific challenges related to contamination or leakage issues owing to their spatial dimension and the possibility of covering the entire population of interest (e.g. all counties within a given state). Despite the peculiarities of aggregate panel data, most of the insights discussed below fully apply to longitudinal microdata on individuals or firms.

---

[1] We emphasize that the term 'data leakage' has multiple meanings and is applied across various domains. In this work, we exclusively refer to the definition established in the ML literature, without any connection to its usage in entirely separate fields, such as cybersecurity—where the term is often used interchangeably with 'data breach' to describe unintentional access to data by unauthorized third parties.

Finally, another strand of literature has adapted and developed ML techniques for causal inference, i.e., to estimate treatment effects rather than predict outcomes (e.g. Wager & Athey, 2018; Chernozhukov et al., 2018), giving rise to the subfield of causal machine learning. Although many of the issues discussed here are also relevant for causal ML techniques when used in conjunction with panel data, our main focus here is on the use of ML and panel data for policy targeting.[2]

The remainder of the paper is arranged as follows. Section 2 discusses the leakage problem. Section 3 provides empirical guidelines for practitioners. Section 4 illustrates the empirical application, while Section 5 concludes.

## 2. The leakage problem

The primary objective of supervised ML is to minimize out-of-sample error when predicting a target outcome based on a set of input features applied to previously unseen data. The standard ML approach involves randomly splitting the original sample into two completely disjoint sets—for example, 80% for training (the training set) and 20% for testing (the testing set). This approach adheres to a 'firewall' principle: none of the data used to generate the prediction function are employed for its evaluation (Mullainathan & Spiess, 2017). The out-of-sample performance of the model on unseen data from the testing set serves as a reliable measure of its 'true' performance on future data.[3] The reliance on randomization works well in classic ML tasks where the data do not have either an explicit temporal or cross-sectional dimension, or can, in any case, be reasonably considered as independent and identically distributed (i.i.d) over both the time and cross-sectional dimensions. However, if applied to panel data, the above standard procedure is wrong, both conceptually and practically, because the i.i.d assumption is violated and this will lead to two different types of data leakage: temporal and cross-sectional leakage. Temporal leakage occurs when information from the future leaks into the past during the training and validation process,[4] whereas cross-sectional leakage refers to information leaks that occur when the same or similar units appears in both the training and testing sets.

---

[2] For econometric work proposing sampling and cross-validation (CV) strategies to prevent data leakage in causal machine learning, see Clark and Polselli (2025) and Furh and Papier (2024).
[3] To address the bias-variance trade-off and prevent overfitting, automatic tuning techniques, such as random k-fold CV, can be applied to the training sample to select optimal hyperparameter values.

The entire point of ML is that the out-of-sample performance constitutes a proxy of the real ability of the model to predict on data that it has *never* encountered before. The key problem with data leakage is that testing set data points are not new to the model. This implies the inflation of the model's out-of-sample performance, which creates the illusion that the ML algorithm can accurately predict the target phenomenon. However, due to data leakage, the algorithm's performance will be substantially poorer when making predictions on genuinely unseen data. Therefore, data leakage can result in misleading policy recommendations.

To see how leakage due to a random training–testing split can occur with panel data, consider, as a possible structure of a longitudinal dataset, the one reported in Figure 1, consisting of panel data with T=7 and N=20, with each unit (e.g. county) belonging to a larger geographical unit (e.g. state, with G=5 in this case). Panel data consist of unit-time observations, where each unit is observed at multiple points in time. In the usual tabular form, the rows of the dataset are therefore unit-time observations. Panel A of Figure 1 illustrates what happens to the data when applying the random split at the unit-time level. Some rows, i.e., unit-time observations, will end up in the training set, whereas others will end up in the testing set. In many instances, the same unit will appear in both the training and testing sets. All the time periods will appear in both the training and testing sets. The ML model will be trained and tuned on the training set, and then it will be employed to predict out-of-sample on the held-out unit-time observations.

At this stage, however, the model will not encounter previously unseen data: it has already 'seen' most units and all time periods belonging to the testing set during its training phase (e.g. in Panel A, county 7 ends up in the training set in all years but year 5, in which it ends up in the testing set). It already 'knows' both most or all of the units and their characteristics—especially if some of the predictors are time-invariant or slowly changing over time—and any temporal trend of the outcome trajectories because it has also been trained on the latest available data. Possibly worse, data at time $t+k$, with $k≥0$, in the training set may be employed to predict the outcome at time $t$ in the testing set. Therefore, with an observation (unit-time) level random split on panel data, there will be both temporal and cross-sectional leakage, leading to a potentially severe overestimation of the ability of the ML model to predict on new data points.

---

[4] As we discuss later, other than through sample splits, temporal leakage can also result from the use of contemporaneous predictors (see Section 3).

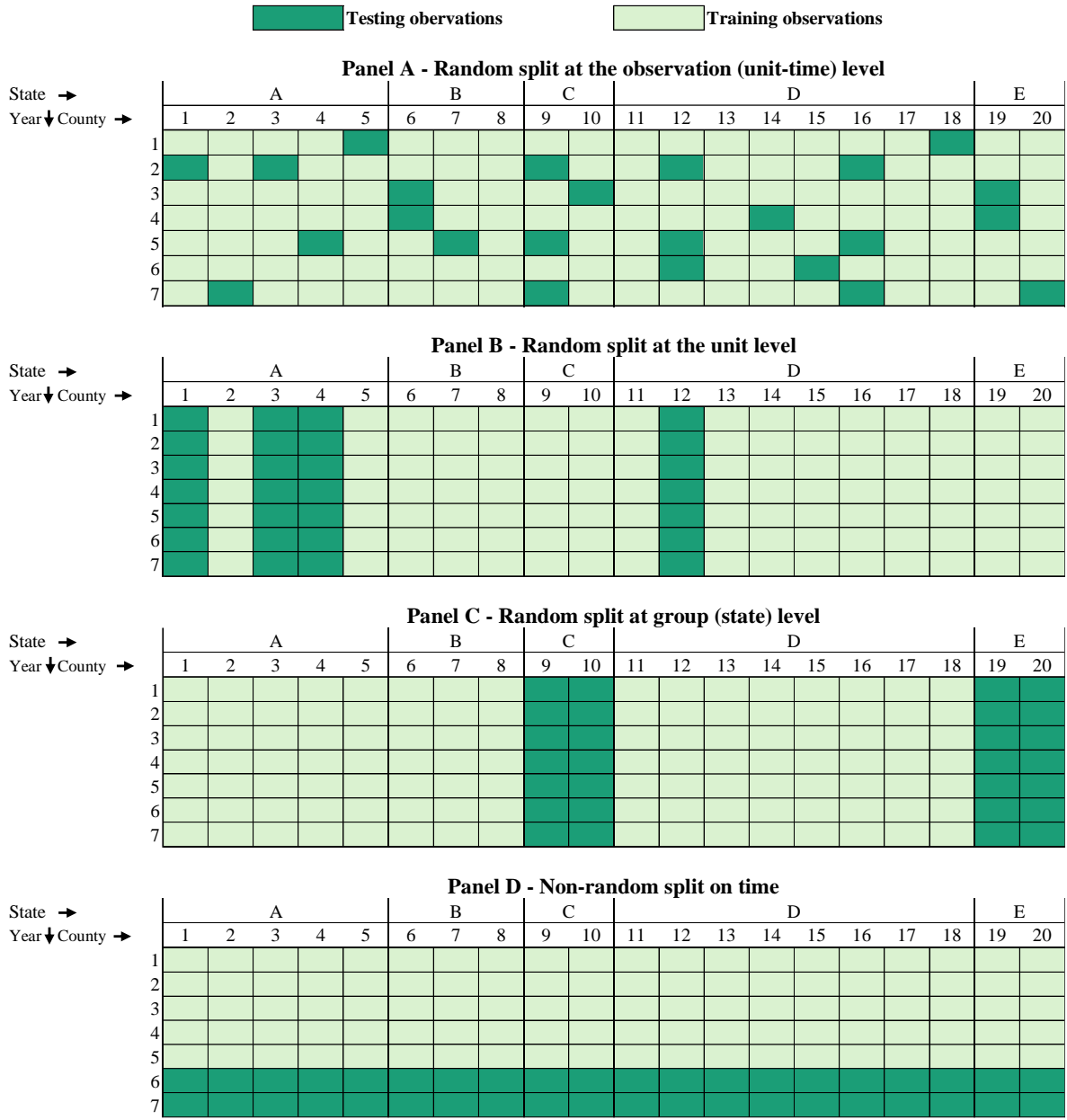

**Figure 1: Training and testing sets under different splitting rules**

*Notes*: This example mirrors a short version of the data from the empirical application described in Section 4, where the time variable corresponds to years, the unit variable to counties, and the group variable to states.

However, alternative solutions can be employed: as Figure 1 shows, one can also split at the unit or group level or, alternatively, at the time level only. Table 1 below reports the four cases depicted in the figure and compares what happens to the observations. With a random split at the unit level, some units will appear only in the training set, whereas the remaining units will feature only in the testing set.[5] All time periods will be present in both sets. This type of split—using an 80/20 rule—is shown in Panel B of Figure 1. This split strategy will result in temporal leakage, specifically in the form of 'trend leakage,' but not cross-sectional leakage. The units that the ML model encounters in the testing set will be previously unseen; however, the algorithm may have already absorbed common shocks or a general trend in the trajectory of the outcome. A variation of this type is the split at the group level, where all the units of some groups appear in the training set, and all the units of other groups feature in the testing set. An example is provided in Panel C of Figure 1, where counties are grouped into states. This approach is usually used to remove residual cross-sectional leakage, as discussed below, but it still suffers from trend leakage. In the fourth case, the researcher can split non-randomly on time: in this case, the first time periods will appear in the training set, whereas the last periods will appear in the testing set (see Panel D of Figure 1 for an example where the first 5 time periods are part of the training set and the last two time periods constitute the testing set).[6] All the units will be present in both sets. In this scenario, there will be cross-sectional leakage but not temporal leakage: the model will already know all the units it will encounter in the testing set, but it has never encountered before data from the latest points in time.[7]

---

[5] In practical implementation, this involves sampling blocks of rows, with the blocks defined by unit groups.

**Table 1: Split strategies for ML with panel data**

| Strategy | Implication | Consequence |
| --- | --- | --- |
| [1] Random split at the observation (unit-time) level | Many units appear in both training and testing sets. All the time periods appear in both the training and testing sets. | Temporal and cross-sectional leakage |
| [2] Random split at the unit level | Some units appear in the training set, while others feature only in the testing set. All time periods are present in both sets. | Temporal (trend) leakage |
| [3] Random split at the group level | All units of some groups appear in the training set, while all units of other groups feature only in the testing set. All time periods are present in both sets. | Temporal (trend) leakage |
| [4] Non-random split on time | Earlier time periods appear only in the training set, while later periods appear exclusively in the testing set. All units are present in both sets. | Cross-sectional leakage |

---

[6] This corresponds to sampling data by the values in the 'Time' column of the dataset.

[7] A fifth, and more complicated, possible splitting criterion is a combination of splitting rules B/C and D in Figure 1, i.e., a rule under which combinations of time periods and units (or groups) appear only either in the training set or the testing set. This splitting rule can prevent both temporal and cross-sectional leakage; however, it may be overly restrictive depending on the research goal and could lead to unsatisfactory predictive performance. A recent paper proposing a CV strategy for panel data resembling this approach is the one by Babii et al. (2023), as further discussed below. Furthermore, in cases where strong cross-sectional and temporal dependencies coexist, two additional potential solutions, complementary to sample-splitting strategies, emerge: (i) explicitly modeling these dependencies through a factor structure, or (ii) leveraging them using more advanced techniques, such as the approach proposed by Chen (2024).

Given that some form of leakage will inevitably always occur when ML is applied to panel data, what is the best possible splitting strategy? To answer this question, we need to take a step back and consider the research goal. We stress that prediction policy problems can be divided into two main—and very different—types:

A. *Cross-sectional prediction policy problems:*

- The rationale for cross-sectional prediction[8] is to address the challenge that arises when data for a specific outcome of interest are available only for a subset of units within a given population. For example, one variable might be collected over time only for certain areas (e.g. large areas). However, both policymakers and researchers aim to comprehensively understand and map the phenomenon across the entire population of units.[9] In this scenario, ML can be applied to units with available outcome information. The model, trained on a subset of these data, can then predict out-of-sample outcomes for the remaining units with observed labels. If the model performs "well", it can be used to predict outcomes for units where all predictor data are available but no outcome data exist. This type of prediction task can be seen as using ML algorithms for missing data imputation rather than as a genuine policy targeting exercise.

  For this cross-sectional prediction policy problem, the most appropriate choice is to split the sample at the unit level (split strategy [2] in Table 1). This approach ensures no cross-sectional contamination, albeit at the expense of possible temporal (trend) leakage. The shared temporal information between the training and testing sets is not problematic in this context because our goal is not to forecast outcomes at future time points to *ex-ante* inform interventions; rather, we aim to map the phenomenon cross-sectionally across units.

B. *Sequential forecasting policy problems:*

- In this case, both the policymaker and the researcher are interested in machine predictions based on historical data that can accurately forecast future outcomes. The final policy goal might be the development of an early warning model or the implementation of preventive policies. Here, the split must rigorously be non-random with respect to time (split strategy [4] in Table 1). The model will be trained and tuned via earlier periods and evaluated on future periods. Operationally, there will be cross-sectional leakage because all units will appear in both the training and testing sets. However, given that the ultimate goal is to produce accurate outcome forecasts for the *same* units at future time points, this cross-sectional leakage is not conceptually or methodologically problematic.

---

[8] We label this type of out-of-sample prediction problem as 'cross-sectional' because, even though we are in a panel setting, its main goal is to produce 'horizontal' predictions of the outcome for other units—hence, cross-sectionally—rather than forecasting outcomes for the same units over time.

[9] This is also known as 'transfer learning' in data-scarce environments.

On the other hand, a random split based on unit-time (split strategy [1] in Table 1) is always problematic, regardless of whether the goal is cross-sectional prediction or forecasting. In this scenario, both types of leakage occur. Which type of leakage will be problematic will depend on the ultimate goal, but regardless of the case, the real-world utility of the model will be overstated, and the machine predictions will be biased, leading to misleading findings. Unfortunately, the discrepancy between evaluated and actual performance on new data will most likely remain concealed until the end of the 'production' stage—i.e., if and when the ML tool is deployed for policy purposes, with all the unintended consequences for the cost, targeting, and effectiveness of said policy.

The preceding discussion is centered around the random split, dividing the original sample into two distinct sets. However, the same principles apply when considering traditional cross-validation (CV) for hyperparameter tuning and model selection on the training set.[10] Let us focus on temporal leakage as our benchmark. When we ignore the temporal dimension of the panel data and perform random k-fold CV (as is automatically, but implicitly, done in most user-friendly ML packages[11]), we end up training the model on *k–1* folds that include future time periods. We subsequently test its performance on the left-out $k^{th}$ fold, which contains past time periods. Unfortunately, this approach does not optimize hyperparameters for forecasting, potentially leading to suboptimal model selection, as the model will not be specifically trained to forecast future observations. These issues related to information leakage across the temporal dimension are not novel and have been known for decades in time series and macroeconomic forecasting analysis. Concepts such as time-series CV (Hyndman & Athanasopoulos, 2018) based on rolling and expanding windows were specifically developed to address these challenges. However, these techniques have only rarely been adapted to specific panel data applications and almost exclusively for macroeconomic and trading forecasting (e.g., Goulet Coulombe, 2024; Goulet Coulombe et al., 2022; Babii et al., 2024; Bluwstein et al., 2023). In this regard, the works by Babii et al. (2023) and Babii et al. (2024) stand out as notable exceptions that propose more general approaches. Babii et al. (2023),combine a block-sampling approach that respects the temporal dimension with a CV strategy that samples units rather than unit-time observations, thus strongly mitigating the risk of data leakage. They also propose using information criteria such as BIC, AIC, and AICs, which are particularly suitable for linear models and are substantially less computationally intensive than CV. Babii et al. (2024), instead, propose a time-series CV strategy to prevent temporal leakage in ML models, which is based on an adaptation of leave-one-out CV with a 'gap' procedure that excludes observations between the training and testing sets, fully decorrelates them, and avoids bias associated with standard k-fold CV. More generally, however, the literature still lacks systematic guidance or an organizing framework for adapting CV to panel data. As noted in a recent review of causal panel data methods by Arkhangelsky and Imbens (2024), the panel dimension introduces additional challenges in implementing CV routines. With the growing application of ML to panel data, we find it crucial to emphasize these concepts.

---

[10] See Kuhn and Johnson (2024) for a detailed discussion on possible CV strategies in predictive settings with temporal and/or spatial autocorrelation in the data. Concerning *causal* ML methods, instead, see the cross-fitting method proposed for the LASSO by Semenova et al. (2023).

[11] There are a few well-known exceptions to random CV, such as the *createTimeSlices* function in the popular R package *caret* (Kuhn, 2008*)*, or the *TimeSeriesSplit* function in the *scikit-learn* Python library (Pedregosa et al., 2011). However, these functions are only suitable for time series, not for panel data. The only routines for panel data CV currently known to us are two new packages developed by Cerqua et al. (2024) for R and Frey (2024) for Python, which can be found here and here.

Furthermore, and distinct from leakage issues, there is also a compelling case for preserving the temporal structure of the data so that the ML models can implicitly account for the time dimension. Supervised ML models (XGBoost, Random Forest, etc.) usually employed in recent literature, in fact, do not explicitly incorporate the temporal (and spatial) dimension of longitudinal data. While there are deep learning algorithms, such as the class of models called Recurrent Neural Networks, that are designed to model and learn sequential data, such as language processing and time series, they are seldom used in conjunction with panel data. This is because, like most time series forecasting techniques, they typically require a substantial number of time periods to obtain accurate predictions. In any case, when applied on tabular data, some supervised ML algorithms, such as tree-based methods, still consistently outperform deep learning techniques (Grinstazjn et al., 2022).

Furthermore, there is also another subtle type of temporal leakage that should be considered when addressing sequential forecasting policy problems, specifically related to the periodic updates of historical data. Indeed, many statistical institutions regularly revise demographic and socio-economic longitudinal data by integrating new data sources, applying enhanced methodologies, and updating relevant classifications (e.g., adopting different industry classifications). Such updates are primarily conducted to improve accuracy, harmonization, and alignment with international statistical standards. Consequently, panel datasets constructed from finalized data may not accurately reflect the actual information available to forecasters in real-time. The standard remedy in the forecasting literature is to rely on real-time historical vintages available at a specific point in time (see Babii et al., 2024;

Ghysels et al., 2018).[12]

Finally, a consideration on cross-sectional prediction problems in scenarios where spatial dimensions play a significant role, such as when working with aggregate panel data. For example, when strong spatial dependence exists among units—as is often the case with economic data (Müller & Watson, 2024)—we encounter a subtle form of contamination. Even if the same unit never appears in both the training and testing sets, if units closely spatially autocorrelated with that unit do appear, spatial leakage occurs. Essentially, the algorithm has not directly "seen" that specific unit, but it "knows" very similar units encountered during training. This might not be a problem if the units for which we are imputing the missing outcome variables share similar characteristics with those used in the training–testing process. However, missing values in panel data are seldom missing at random. If this is the case, cross-sectional leakage may potentially lead to an inflation of the performance of the ML algorithms with respect to the true untestable performance on the observations with missing outcome data. To mitigate this issue, consider applying a random split not at the individual unit level but at a higher level of clustering (split strategy [3] in Table 1). For example, if the unit is a county, split the data at the state level. By stratifying the sample in this way, we reduce the risk of spatial leakage.[13]

While many of the insights discussed in this section may seem intuitive, they are rarely acknowledged or systematically addressed in empirical practice. For instance, among all the nine studies cited in the introductory paragraph, none explicitly mentions the term 'data leakage.' Furthermore, with the sole exception of Bluwstein et al. (2023), none provides a detailed account of how the ML pipeline mitigates potential leaks that could lead to inflated out-of-sample performance estimates.

Having highlighted these common yet often overlooked pitfalls, the next section provides a set of practically relevant recommendations for researchers dealing with these challenges in applied settings.

---

[12] In the empirical application illustrated in Section 4, the revision of historical data might lead to temporal data leakage if data from time t are used to update data at times t-1, t-2, and so forth. However, this is not an issue in our application, which relies on data from the U.S. Bureau of Economic Analysis, which is not subject to such leakage. In fact, although the U.S. Bureau of Economic Analysis periodically updates historical data, changes in values are not driven by future data but rather by: i) improved classification and measurement of real estate investment trusts, regulated investment companies, housing services, and own-account software investment; and ii) enhancements to the estimation methods used for preparing subnational statistics. For more details, see here.

[13] This is the equivalent solution to what is done in causal inference to address potential violations of the stable-unit-treatment-value assumption in contexts with spatial spillovers.

## 3. Practical guidelines

Table 2 below summarizes our recommendations for practitioners. Note that these guidelines pertain to both classification and regression problems. First and foremost, the researcher should clarify the research goal at the outset: which type of prediction policy problem you are working on? Are you interested in cross-sectional prediction or forecasting? On the basis of the answer to this preliminary question, the entire subsequent ML panel pipeline is designed accordingly.

An important step that precedes the sample split and CV stages, which we did not discuss earlier, concerns the choices governing the selection of predictors included in the dataset. This choice also depends on the overall goal of the project and is particularly important when the objective is forecasting for *ex-ante* policy targeting. If the researcher is interested in forecasting, it is crucial that only lagged (or time-invariant) predictors are included in the ML model. In this case, including predictors contemporaneous with the outcome—or even subsequent to it—would be both methodologically and practically incorrect. Methodologically, it would result in simultaneity issues, which can be seen as a form of temporal leakage originating from the predictors, as the ML model would learn any distribution shift or structural break at time $t$ that is common to both the outcome and the predictors. In addition, the predictors themselves might be affected by the predicted event. It is indeed well-established in forecasting practice that, to forecast future values of a variable, only information available at the time the forecast is made can be used and that the forecasting ability of a model must be evaluated by generating forecasts over some past period (with known outcomes) only using data known at each forecast origin (Petropoulos et al., 2022). Moreover, it is generally beneficial to include lagged outcome values as additional predictors to increase the forecasting accuracy of the ML algorithm.

On the practical side, if the goal is to provide policymakers with machine predictions that can anticipate the unfolding of a given phenomenon, then the applicability of those predictions should depend on data promptly available to the policymaker before the outcome materializes. Including contemporaneous predictors would mean waiting for those data to be collected, effectively turning the forecasting problem into a retrospective one by the time the predictor data become available. Finally, no direct derivations or mechanical transformations of the outcome should be included as predictors (e.g. the use of contemporaneous income levels to predict income growth).

**Table 2: Do's and Don'ts with ML analysis on panel data**

| ML Pipeline Step | *Do's* | *Don'ts* |
|---|---|---|
| **1. Research design** | Clarify your research goal at the outset: are you interested in a cross-sectional prediction problem or a sequential forecasting problem? | Skip this stage and start the empirical analysis without having your goal in mind *ex-ante*. |
| **2. Dataset building** | For cross-sectional prediction, avoid including variables that are direct derivations or transformations of the outcome. For example, if you are predicting GDP, do not use the logarithm of GDP as a predictor.<br><br>If you are interested in sequential forecasting, all predictors should be measured at least at $t-1$. Include lagged values of the outcome as additional predictors.[14] As above, do not include variables that are a direct derivation or transformation of the outcome. | Include both contemporaneous and lagged predictors, regardless of the problem's nature, or variables that are a direct derivation of the outcome. |
| **3. Sample split** | For cross-sectional prediction, split randomly at the unit level. If you suspect residual cross-sectional (e.g. spatial) leakage, consider stratified sampling at a higher level of aggregation.<br><br>For sequential forecasting, split non-randomly on time. | Apply a random split at the observation (unit-time) level. |
| **4. Cross-validation** | If your goal is cross-sectional prediction, apply stratified CV at the unit or group level on the training set.<br><br>If interested in forecasting, apply temporal CV at the time level on the training set (using either an expanding or rolling window approach).[15] | Apply random k-fold CV. |
| **5. Out-of-sample testing** | Evaluate the out-of-sample performance of the model on truly previously unseen data. | Test the out-of-sample performance of the model on data it has already encountered before. |

In the next section, we provide an empirical illustration that demonstrates how to detect and quantify the degree of data leakage associated with the uncritical application of ML to panel data.

## 4. Empirical Application

To illustrate the practical implications of the issues discussed above, we carry out a comprehensive analysis on a panel dataset from the U.S. Specifically, the analysis is conducted at the county level for all U.S. states. We created a balanced panel of 3,058 counties (out of the 3,143 existing counties) from 2000-2019.[16] We collected a variety of variables to analyze economic performance across the U.S. The list of variables and their sources is reported in Table A.1 in Online Appendix A, while the descriptive statistics are in Table A.2 in Online Appendix A. The notebook replicating the analysis presented here is accessible on the GitHub repository page accompanying this paper. Versions of these data have been used in many applied studies—for example, see Foote et al. (2019) and Mayda et al. (2022).

---

[14] Ideally, the choice of the optimal lag structure should also be cross-validated and based on a grid search.

[15] As most existing ready-to-use package routines automatically implement random k-fold CV, this will likely involve a preliminary pre-processing step in which researchers manually prepare the temporally-ordered folds to implement a panel version of cross-validation which carefully preserves the temporal ordering of the data.

The overall objective of our analysis is to predict or forecast economic outcomes. We carry out two separate analyses, a classification task and a regression task, to quantify and document data leakage across both types of prediction settings. There are two outcomes:

- For the regression task, we use personal income per capita;
- For the classification task, we use a recession dummy that indicates whether a county experienced a decrease in personal income per capita in a given year.[17]

There are two different prediction policy problems, as outlined in our previous discussion:

i) sequential forecasting;
ii) cross-sectional prediction.

Additionally, since we discussed above that temporal leakage in forecasting problems can be particularly severe if there are unforeseen changes in the relationships among variables, we also explore the forecasting problem with a special focus on a single year of our panel dataset—2009—during which the U.S. experienced a drop in per capita income due to the Great Recession.

This means that we have a total of six different predictive problems to solve:

1) Forecasting a binary outcome;
2) Forecasting a continuous outcome;
3) Predicting cross-sectionally a binary outcome;
4) Predicting cross-sectionally a continuous outcome;
5) Forecasting a binary outcome in a particular year (2009);
6) Forecasting a continuous outcome in a particular year (2009).

---

[16] We lose a few counties due to boundary changes over the period under analysis and a few others because they lack data for at least one of the variables considered.

[17] Although a recession is generally considered to occur when there is a decline in real GDP for at least two consecutive quarters, we cannot adopt such definition for two reasons: i) our panel is at the year level; ii) the U.S. Bureau of Economic Analysis provides data on GDP at the county level only from 2017 onward.

To solve each of these problems, we use the same set of raw predictors. This ensures that any differences in performance that emerge can only be due to different partitioning of the data and the selection of the raw predictors.

While this application is intended as an illustration of the presence and severity of temporal and cross-sectional leakage in ML with panel data, rather than as a topic of specific interest in itself, this setting is clearly relevant for forecasting policy problems, as it uses socioeconomic data from a widely studied high-income country, which could, for example, be employed to forecast areas that might soon enter a recession. Such analyses have self-evident policy implications and relevance for both the U.S. and the global economy. Predicting income cross-sectionally is of less immediate interest in itself; however, we use the same outcome for the cross-sectional exercise to ensure consistency and comparability of the magnitude of different leakage types. In similar contexts, this type of data could be used cross-sectionally, for instance, to impute missing subnational inflation data that are collected only for larger or economically significant areas, under the implicit assumption that the data-generating process is consistent across areas with and without inflation data.

For each predictive task (classification and regression), we run several different models featuring various combinations of the following parameters:

1. **Use of contemporaneous predictors**: In forecasting problems, this is one of the main types of data leakage. We alternatively include and remove contemporaneous predictors for the different models. For example, when we include contemporaneous predictors, we use unemployment in 2014 to predict income per capita in 2014. However, we never include contemporaneous predictors that are a direct derivation of the outcome variables (e.g. we do not use log income per capita in 2014 as a predictor in the classification task for the same year).
2. **Sample split strategy**: We compare all the split strategies illustrated in Figure 1 and Table 1: [1] random split: a random split of the dataset at the county-year level; [2] county split: we randomly split at the county level, assigning counties to either the training or the testing set; [3] state split: we randomly split at the state level, and then assign their counties to either the training or the testing set, to reduce potential spatial leakage; [4] time split: a non-random split at the year level.[18]
3. **Inclusion of lagged outcomes:** This criterion is only used to assess how performance varies across models with and without outcome lags included**.**

---

[18] For the non-random split at the year level we assign all the observations in the years 2016, 2017, 2018 and 2019 to the testing set.

4. **Adjustment of the testing set size:** without adjustments, the different choices in the above criteria result in different testing set sizes for the separate models. We alternatively adjust or not adjust the testing set size of the different models and run both the adjusted and unadjusted versions of the models to ensure an adequate comparative performance assessment.
5. **Algorithm:** We employ two different ML algorithms: extreme gradient boosting (XGBoost) and Random Forest. These algorithms are among the most popular ML techniques used by applied researchers. Refer to Online Appendix B for a short description of these models. For comparability, we also run simpler models, namely, a Logit model for the classification problem and ordinary least squares (OLS) for the regression task.

Note that among the above-described parameters, only choices involving points 1) and 2) can be sources of data leakage, with 1) being a source of (temporal) leakage only in forecasting problems. Considering the six different prediction problems and the five modeling criteria described above, we run 480 different models. Owing to the large number of models and to ensure better comparability across different configurations, we use default settings for the hyperparameter values of the ML algorithms without cross-validating them. However, remember that when CV is performed, it should follow the guidelines provided in Section 3. We start by discussing temporal leakage in the forecasting problems, and next, we move to cross-sectional leakage.

### 4.1 Temporal leakage (Predictive problems 1, 2, 5, and 6)

Figure 2 provides a comprehensive visualization of the performance of various model configurations for classification (Panel A) and regression tasks (Panel B) across all time periods, where the goal is to forecast an outcome over time and there is a risk of temporal leakage. The full results for each model are also reported in Tables A.3 and A.4 of the Online Appendix.

In our context, temporal leakage can arise from two factors: (a) when contemporaneous predictors are included, and (b) when the dataset is not split based on time. In each plot, a squared marker (□) represents the performance of a model (as measured by the metric reported on the y-axis, either AUC or MSE). For each model family (e.g., Random Forest, XGBoost, Logit/OLS), the different combinations are ordered along the x-axis from the worst- to best-performing. The table of colored squares at the bottom of each plot provides detailed information on the specific combination used in each model. For instance, in the left plot of Panel A—Figure 2, we focus on the Random Forest model (dark-blue marker line) and observe that performance improves (the slope of the line is positive) when the model is affected by temporal leakage. Specifically, this occurs when the model is built on a random split of the data and includes contemporaneous predictors, as indicated by the dark-colored squares in the table at the bottom of the plot.

As we can see, in models with temporal leakage, the performance, as measured by the Area under the Curve (AUC), is significantly higher than that of non-leaked models. For example, in Figure 2 (classification – Panel A), when a Random Forest model is employed, the average AUC among the leaked models is 0.759, whereas the average among non-leaked models is 0.708, a difference of 0.051 points.[19] This gap represents a considerable difference, as the AUC ranges from 0 to 1, with 0.5 being as good as random guessing. Similar results are observed with XGBoost (center image in Figure 2 Panel A) and Logit models (right image in Figure 2 Panel A). Analogous insights, with even greater degrees of leakage, apply to the regression task (see Figure 2 Panel B), where performance is measured by Mean Squared Error (MSE). The MSE of the leaked models is substantially lower. For example, in the case of XGBoost, the leakage ratio—expressed as the ratio between the difference in the MSEs of the leaked and non-leaked models over the MSEs of the non-leaked models—is greater than 17%. This magnitude of difference among forecasting models is generally considered very large in the forecasting literature (Armstrong, 2001).

Figure 3 focuses on the year of the Great Recession (2009). The difference in performance due to temporal leakage can be significantly greater when an unexpected shock occurs. For instance, the AUC for the classification task with the Random Forest model drops from 0.692 for the leaked models to 0.442 for the non-leaked ones (Figure 3 - Panel A). In contrast, the leakage ratio for the regression problem (Figure 3 - Panel B) remains similar to that for the models run across all periods. This is not surprising since the decrease in personal income in the recession year was not large enough to entail significant differences in the value of the continuous outcome variable. These impressive differences for the classification task in 2009 underscore that inflated performance due to temporal leakage is particularly harmful when trying to forecast otherwise difficult-to-anticipate events. In fact, structural breaks, such as the Great Recession, are known in the ML field as distribution shifts. A distribution shift occurs when the training data distribution differs from the data distribution the model encounters during testing.[20] Predictive accuracy typically declines with the magnitude of the distribution shift between the target and testing data (Dell, 2024), leading, in many cases, to a sharp drop in out-of-sample performance because the input-to-output relationships and patterns the model learned during training no longer hold true in the new environment. By definition, distribution shifts are impossible to predict, as they imply a fundamental and unforecastable change in the data-generating process. In our context, even a well-performing ML model cannot anticipate the impact of the Great Recession on economic outcomes when trained and tuned only on past information. This type of event thus represents a decisive litmus test to highlight the problem with leakage issues: if the model performs well on post-break data and the prediction error is small, something is probably wrong. Likely, it is not because the model can magically predict the future but rather because some form of leakage—information from the future—has sneaked into the model, either via observations or predictors (or both). Failing to realize this means severely overestimating the power of ML models.[21]

---

[19] Regarding statistical significance, we are not aware of statistical tests *à la* DeLong et al. (1988) that apply to groups of ROCs, rather than individual comparisons among ROCs, especially when the models are run on different data. Although uncertainty quantification is theoretically possible, it is very computationally demanding in practice due to the number of models involved. Similar considerations apply to the subsequent comparisons below.

[20] Here, the reason is an unexpected recession, but there may be various reasons for distribution shifts, such as a discontinuity in the mode of data collection, or when the type of data on which the model is tested in the production stage differs fundamentally from the data source on which it was trained. For instance, a switch from face-to-face to phone-based interviews (which occurred for most surveys during the COVID-19 pandemic) can structurally alter the nature of the data. Similarly, a domain shift can occur when deep learning algorithms trained on modern web texts are tested on fundamentally different data from ancient texts or historical archives (Dell, 2024).

---

[21] More generally, one should be skeptical *a priori* of extremely good out-of-sample performances in predicting or forecasting complex socioeconomic outcomes, which are partially characterized by inherently unpredictable idiosyncratic patterns that make the irreducible error substantially higher than in most standard ML applications.

**Figure 2: Temporal leakage in the forecasting problem (all years)**

**Panel A - Classification**

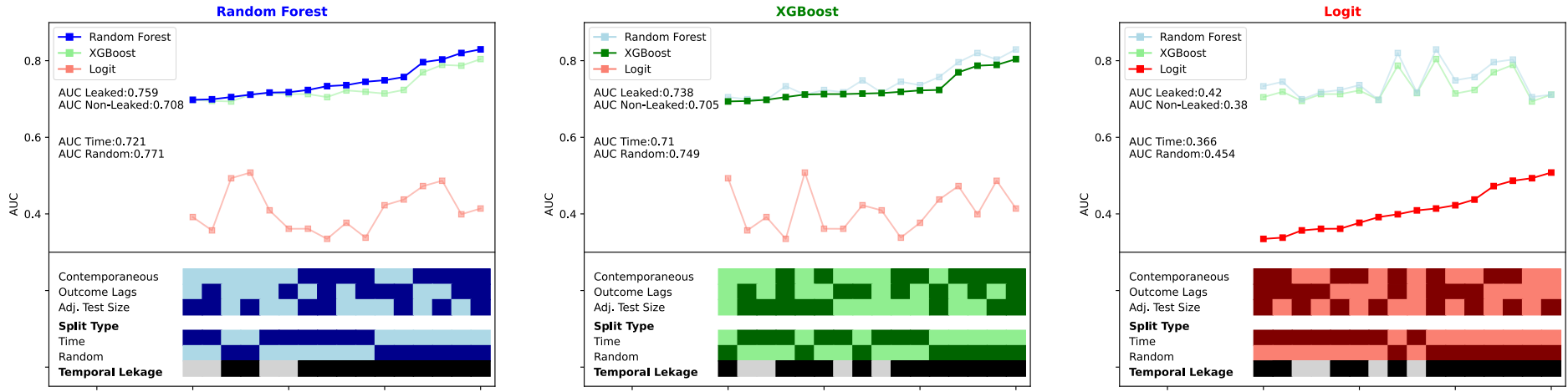

**Panel B - Regression**

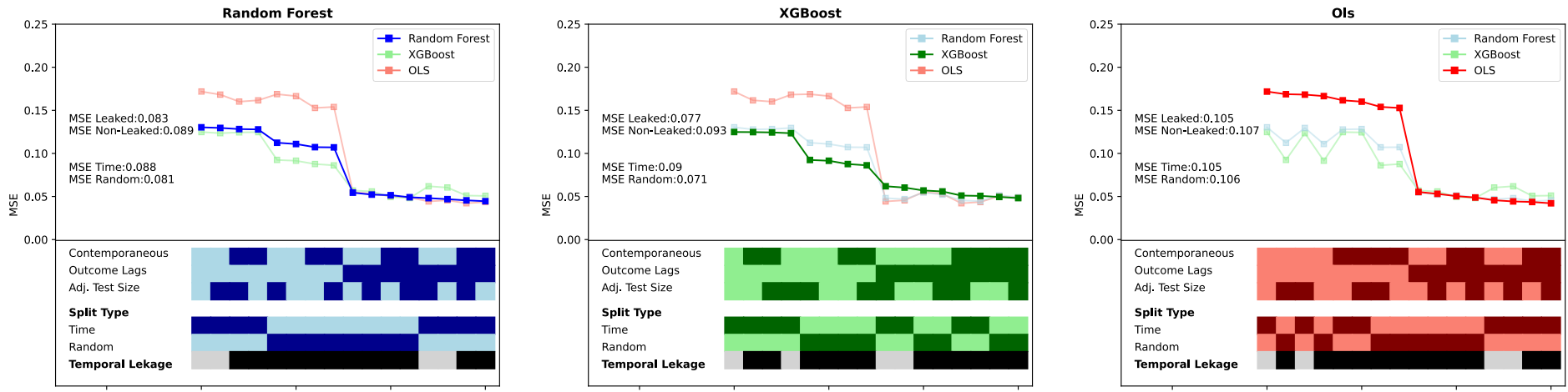

*Notes:* Each plot shows the performance and characteristics of different models, where we forecast a binary dummy for recession (Panel A) or the log of income per capita (Panel B) using a set of predictors (as described in Online Appendix A). We focus on one family of models at a time (left = Random Forest, center = XGBoost, right = OLS/Logit) but keep other models in the background (light-colored lines) for comparison. Each square marker (e.g., □) represents a unique model, ordered from the worst- to best-performing along the x-axis based on the chosen metric (y-axis). In the table of squares at the bottom of each plot, we report the characteristics that describe the combination of each model. The different combinations are highlighted by colored rectangles below each marker, where darker colors indicate the activation of that parameter. The bottom black/grey bar summarizes whether the model is temporally leaked. Specifically, a black rectangle represents a temporally leaked model that includes at least one of the following: contemporaneous variables or a non-time-based split. The full results are reported in Tables A.3 and A.4 of Online Appendix A.

**Figure 3: Temporal leakage in the forecasting problem (focus on the Great Recession, year 2009)**

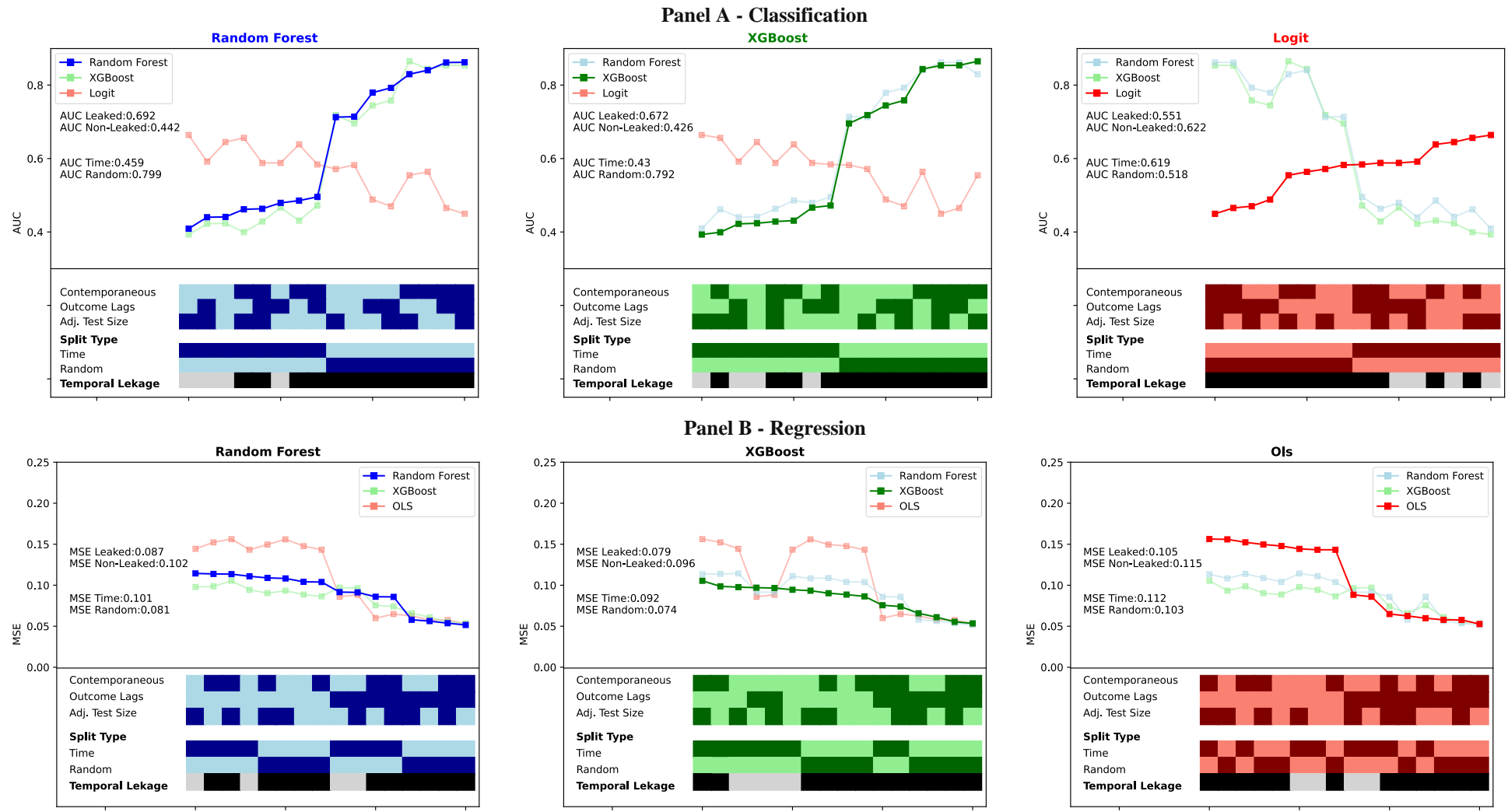

*Notes*: Each plot shows the performance and characteristics of different models, where we forecast a binary dummy for recession (Panel A) or the log of income per capita (Panel B) using a set of predictors (as described in Online Appendix A). We focus on one family of models at a time (left = Random Forest, center = XGBoost, right = OLS/Logit) but keep other models in the background (light-colored lines) for comparison. Each square marker (e.g., □) represents a unique model, ordered from the worst- to best-performing along the x-axis based on the chosen metric (y-axis). In the table of squares at the bottom of each plot, we report the characteristics that describe the combination of each model. The different combinations are highlighted by colored rectangles below each marker, where darker colors indicate the activation of that parameter. The bottom black/grey bar summarizes whether the model is temporally leaked. Specifically, a black rectangle represents a temporally leaked model that includes at least one of the following: contemporaneous variables or a non-time-based split. The full results are reported in Tables A.5 and A.6 of Online Appendix A.

### 4.2 Cross-Sectional Leakage (Predictive problems 3 and 4)

We now turn to cross-sectional leakage, which in our case, since we use aggregate panel data, takes the form of spatial leakage—a specific type of cross-sectional leakage. Spatial leakage occurs because the same or similar (spatially autocorrelated) units appear in both the training and testing sets. To assess the varying intensity of this type of leakage, we analyzed how the performance of the models changed with different splitting methods. The benchmark leaked model is trained and tested based on a random split, where all units (i.e., counties) are likely to appear in both the training and testing sets. For the county split (i.e., unit-level split), counties in the training set do not appear in the testing set. For the state split, we ensure that all the counties in the testing set are not in the same state as any county in the training set. In other words, in the county split, the model predicts the outcome of a county that it has never seen before, whereas in the state split, the model predicts the outcome of a county from a state whose counties the model has never seen before.

The results are shown in Figure 4 for the classification (Panel A) and regression tasks (Panel B). As we can see, in all the models except Logit[22], all the leaked models dominate the non-leaked ones, indicating the presence of spatial leakage (as the split strategy is the only major source of difference in the data employed by the different models), although the differences are somewhat less pronounced than those of temporal leakage. Interestingly, the worst performances for the ML models are observed when using state splits, suggesting the presence of residual spatial leakage at the county level due to spatially autocorrelated nearby counties.

In summary, our empirical application shows that, in a typical panel dataset, the data leakage issues discussed in this paper tend to lead to significant overestimation of the performance of ML models across all the predictive problems considered. We are aware that these issues might not always be as severe as in this case study, since the extent of leakage and the degree of out-of-sample performance overestimation will vary depending on the application. However, even when leakage is proven to be minimal, it remains conceptually and methodologically inappropriate to disregard the unique characteristics of panel data in such applications.

---

[22] Note that the Logit model, unlike the other two ML algorithms, has difficulty handling the unbalanced distribution of the outcome variable (recession dummy) for the classification task. In these cases, performance metrics like the AUC can be uninformative. To understand this problem with Logit, check Table A.3, where we also report detailed results on sensitivity and specificity across the different classification models. Consequently, the Logit results for classification should be interpreted with a grain of salt.

## Figure 4: Cross-sectional leakage (all years)

### Panel A - Classification

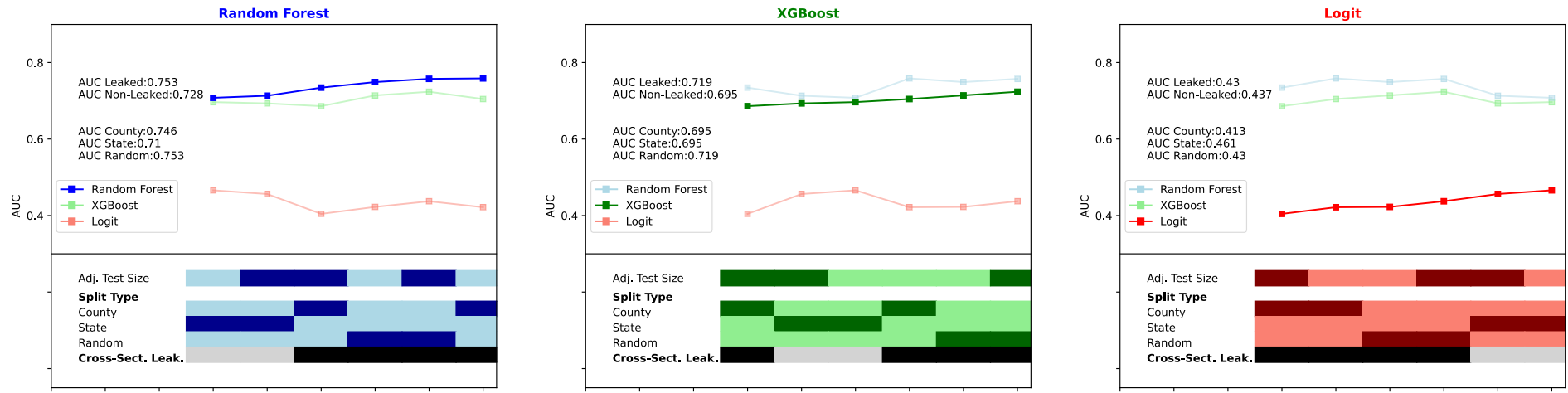

### Panel B - Regression

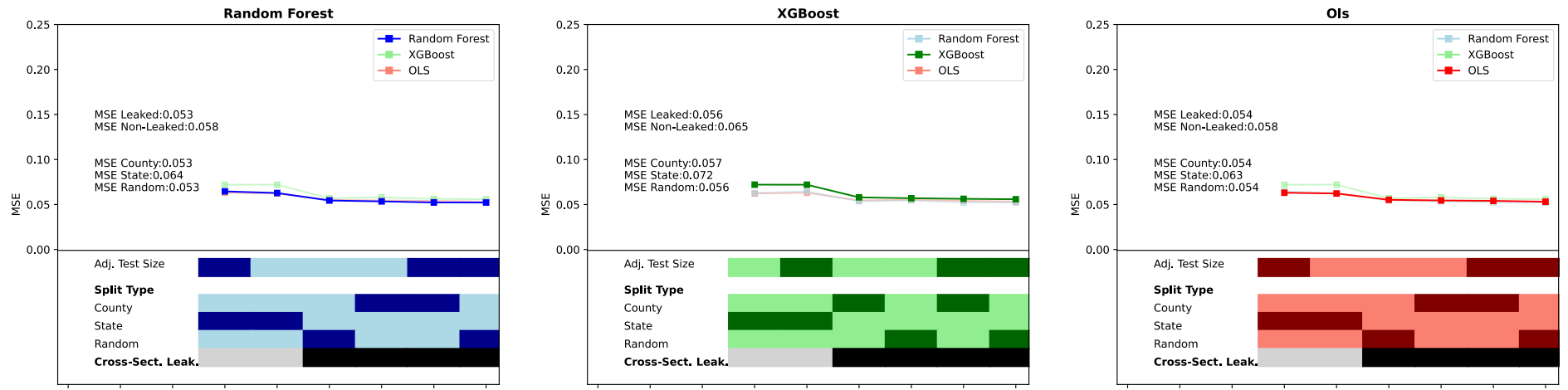

*Notes*: Each plot illustrates the performance and characteristics of different models, where we forecast either a binary dummy for recession (Panel A) or the logarithm of income per capita (Panel B) using a set of predictors (as described in Online Appendix A). We focus on one family of models at a time (left = Random Forest, center = XGBoost, right = OLS), while keeping other models in the background (represented by light-colored lines) for comparison. Each square marker represents a unique model, ordered along the x-axis from the worst- to best-performing based on the chosen metric (y-axis). The table of squares at the bottom of each plot describes the characteristics of the combination used in each model. The different combinations are highlighted with colored rectangles below each marker, where darker colors indicate the activation of a specific parameter. The bottom black/grey bar summarizes the presence and severity of cross-sectional leakage (darker color → higher leakage). Specifically, a black rectangle denotes a spatially leaked model with a training/test set split not determined at the state level. Full results are provided in Tables A.7 and A.8 of Appendix A.

# 5. Conclusions

Data leakage, i.e., the unintended use of information during model training and validation that would not be available at the prediction stage, in the use of ML is increasingly recognized as a fundamental challenge across many scientific fields (Bernett et al., 2024; Kapoor & Narayanan, 2023; Rosenblatt et al., 2024). No such widespread awareness seems to have emerged in the social sciences, especially concerning longitudinal data. We argue that more effort and critical thinking should be devoted to the preliminary design of ML analysis on panel data to avoid data leakage issues that might bias out-of-sample performance upward and unintentionally mislead policymakers into overestimating the power and applicability of ML techniques. We caution practitioners against simply importing traditional ML practices into their research domain without taking care of the underlying analytical and practical implications. We further recommend that, as a best practice, every applied ML paper should clearly detail the entire ML pipeline and include a transparent discussion explicitly describing the steps taken to address and prevent data leakage issues. Given the increasing adoption of ML tools in economics and other social sciences and the growing availability of panel data, these issues deserve attention.

**References**

Antulov-Fantulin, N., Lagravinese, R., & Resce, G. (2021). Predicting bankruptcy of local government: A machine learning approach. *Journal of Economic Behavior & Organization*, *183*, 681-699.

Apicella, A., Isgrò, F., & Prevete, R. (2024). Don't Push the Button! Exploring Data Leakage Risks in Machine Learning and Transfer Learning. *arXiv preprint arXiv:2401.13796*.

Arkhangelsky, D., & Imbens, G. (2024). Causal models for longitudinal and panel data: A survey. *The Econometrics Journal*, utae014.

Armstrong, J. S. (2001). *Combining forecasts*. In *Principles of Forecasting*, pp. 417-439. Springer U.S.

Ash, E., Galletta, S., and Giommoni, T. (2024). A machine learning approach to analyze and support anti-corruption policy. *American Economic Journal: Economic Policy,* forthcoming.

Athey, S., Keleher, N., & Spiess, J. (2025). Machine learning who to nudge: causal vs predictive targeting in a field experiment on student financial aid renewal. *Journal of Econometrics*, 105945.

Babii, A., Ball, R. T., Ghysels, E., & Striaukas, J. (2023). Machine learning panel data regressions with heavy-tailed dependent data: Theory and application. *Journal of Econometrics*, *237*(2), 105315.

Babii, A., Ball, R. T., Ghysels, E., & Striaukas, J. (2024). Panel data nowcasting: The case of price–earnings ratios. *Journal of Applied Econometrics*, *39*(2), 292-307.

Babii, A., Ghysels, E., & Striaukas, J. (2024). 10. Econometrics of machine learning methods in economic forecasting. *Handbook of Research Methods and Applications in Macroeconomic Forecasting*, 246.

Battaglini, M., Guiso, L., Lacava, C., Miller, D. L., & Patacchini, E. (2024). Refining public policies with machine learning: The case of tax auditing. *Journal of Econometrics*, available online 23 September 2024.Bernett, J., Blumenthal, D. B., Grimm, D. G., Haselbeck, F., Joeres, R., Kalinina, O. V., & List, M. (2024). Guiding questions to avoid data leakage in biological machine learning applications. *Nature Methods*, *21*(8), 1444-1453.

Bluwstein, K., Buckmann, M., Joseph, A., Kapadia, S., & Şimşek, Ö. (2023). Credit growth, the yield curve and financial crisis prediction: Evidence from a machine learning approach. *Journal of International Economics*, *145*, 103773.

Boss, K., Groeger, A., Heidland, T., Krueger, F., & Zheng, C. (2024) Forecasting bilateral asylum seeker flows with high-dimensional data and machine learning techniques. *Journal of Economic Geography*, lbae023.

Cakici, N., Fieberg, C., Neumaier, T., Poddig, T., & Zaremba, A. (2024). Pockets of Predictability: A Replication. *Journal of Finance,* forthcoming.

Cerqua, A., Letta, M., & Menchetti, F. (2024). Causal inference and policy evaluation without a control group. arXiv preprint:2312.05858v2.

Chen, K. (2024). Inference in High-Dimensional Panel Models: Two-Way Dependence and Unobserved Heterogeneity. *Michigan State Economics Working Paper*.

Chernozhukov, V., Chetverikov, D., Demirer, M., Duflo, E., Hansen, C., Newey, W., & Robins, J. (2018). Double/debiased machine learning for treatment and structural parameters. *Econometrics Journal*, 21(1), C1-C68.

Christensen, P., Francisco, P., Myers, E., Shao, H., & Souza, M. (2024). Energy efficiency can deliver for climate policy: Evidence from machine learning-based targeting. *Journal of Public Economics*, 234, 105098.

Clarke, P., & Polselli, A. (2025). Double machine learning for static panel models with fixed effects. *Econometrics Journal*, forthcoming.

de Blasio, G., D'Ignazio, A., Letta, M. (2022). Gotham city. Predicting 'corrupted' municipalities with machine learning. *Technological Forecasting and Social Change*, 184, 122016.

Dell, M. (2024). Deep learning for economists. *National Bureau of Economic Research (NBER) Working Paper*, No.32768.

DeLong, E. R., DeLong, D. M., & Clarke-Pearson, D. L. (1988). Comparing the areas under two or more correlated receiver operating characteristic curves: a nonparametric approach. *Biometrics*, 837-845.

Foote, A., Grosz, M., & Stevens, A. (2019). Locate your nearest exit: mass layoffs and local labor market response. *ILR Review*, 72(1), 101-126.

Frey, E. (2024). How to cross-validate your panel data in Python. *Towards Data Science*. Available at : https://github.com/4Freye/panelsplit.

Fuhr, J., & Papies, D. (2024). Double Machine Learning meets Panel Data--Promises, Pitfalls, and Potential Solutions. *arXiv preprint arXiv:2409.01266*.

Ghysels, E., C. Horan, and E. Moench (2018). Forecasting through the rearview mirror: Data revisions and bond return predictability. *The Review of Financial Studies*, 31(2), 678–714.

Grinsztajn, L., Oyallon, E., & Varoquaux, G. (2022). Why do tree-based models still outperform deep learning on typical tabular data?. *Advances in Neural Information Processing Systems*, 35, 507-520.

Goulet Coulombe, P. (2024). The macroeconomy as a random forest. *Journal of Applied Econometrics*, 39(3), 401-421.

Goulet Coulombe, P., Leroux, M., Stevanovic, D., & Surprenant, S. (2022). How is machine learning useful for macroeconomic forecasting?. *Journal of Applied Econometrics*, 37(5), 920-964.24

Hastie, T., Tibshirani, R., & Friedman, J. H. (2009). *The elements of statistical learning: data mining, inference, and prediction*. New York, Springer.

Jarvis, S., Deschenes, O., & Jha, A. (2022). The private and external costs of Germany's nuclear phase-out. *Journal of the European Economic Association*, *20*(3), 1311-1346.

Hyndman, R. J., & Athanasopoulos, G. (2018). *Forecasting: Principles and Practice*. OTexts.

Kapoor, S., & Narayanan, A. (2023). Leakage and the reproducibility crisis in machine-learning-based science. *Patterns*, *4*(9).

Kaufman, S., Rosset, S., Perlich, C., & Stitelman, O. (2012). Leakage in data mining: Formulation, detection, and avoidance. *ACM Transactions on Knowledge Discovery from Data (TKDD)*, *6*(4), 1-21.

Kleinberg, J., Ludwig, J., Mullainathan, S., & Obermeyer, Z. (2015). Prediction policy problems. *American Economic Review*, *105*(5), 491-495.

Kuhn, M. (2008). Building predictive models in R using the caret package. *Journal of Statistical Software*, *28*, 1-26.

Kuhn, M., & Johnson, K. (2024). Applied Machine Learning for Tabular Data. Available at: https://aml4td.org/.

Mayda, A. M., Peri, G., & Steingress, J. (2022). The political impact of immigration: evidence from the United States. *American Economic Journal: Applied Economics*, *14*(1), 358-389.

Messeri, L., & Crockett, M. J. (2024). Artificial intelligence and illusions of understanding in scientific research. *Nature*, *627*(8002), 49-58.

Mullainathan, S., & Spiess, J. (2017). Machine learning: an applied econometric approach. *Journal of Economic Perspectives*, *31*(2), 87-106.

Müller, U. K., & Watson, M. W. (2024). Spatial unit roots and spurious regression. *Econometrica*, *92*(5), 1661-1695.

Pedregosa, F., Varoquaux, G., Gramfort, A., Michel, V., Thirion, B., Grisel, O., ... & Duchesnay, É. (2011). Scikit-learn: Machine learning in Python. *the Journal of Machine Learning Research*, *12*, 2825-2830.

Petropoulos, F., Apiletti, D., Assimakopoulos, V., Babai, M. Z., Barrow, D. K., Taieb, S. B., ... & Ziel, F. (2022). Forecasting: theory and practice. *International Journal of Forecasting*, *38*(3), 705-871.

Rosenblatt, M., Tejavibulya, L., Jiang, R., Noble, S., & Scheinost, D. (2024). Data leakage inflates prediction performance in connectome-based machine learning models. *Nature Communications*, *15*(1), 1829.

Sansone, D., & Zhu, A. (2023). Using machine learning to create an early warning system for welfare recipients. *Oxford Bulletin of Economics and Statistics, 85*, 959-992.

Semenova, V., Goldman, M., Chernozhukov, V., & Taddy, M. (2023). Inference on heterogeneous treatment effects in high-dimensional dynamic panels under weak dependence. *Quantitative Economics*, *14*(2), 471-510.

Wager, S., & Athey, S. (2018). Estimation and inference of heterogeneous treatment effects using random forests. *Journal of the American Statistical Association*, *113*(523), 1228-12.

**Online Appendix - Supplementary application information**

**A. Data and full results**

The analysis is conducted at county level for all U.S. states. We have created a balanced panel of 3,058 counties (out of the 3,143 existing counties) from 2000 to 2019. We have collected a variety of variables to analyze economic performance across the U.S.. The list of variables and their sources is reported in Table A.1, while the descriptive statistics are in Table A.2. As described in the main text, there are two dependent variables:

i) one for the regression problem, i.e., personal income per capita growth rate measured as the annual percentage change in personal income per capita[23];

ii) one for the classification problem, namely, a recession dummy that indicates whether a county experienced a drop in personal income per capita in a given year.

Economic variables including personal income per capita, average wage, percentage of income from unemployment benefits and workplace employment rate (number of employees divided by the resident population from 18 to 65 years old) are all sourced from the U.S. Bureau of Economic Analysis (BEA). The rest of the predictors are instead sourced from U.S. Census. Demographic information encompasses total population, the percentage of the population under 18 and over 65, as well as detailed breakdowns by gender and racial composition (White, Black, Hispanic, and Asian percentages of the population). Additionally, it includes birth and death rates per 1,000 inhabitants. Lastly, also the mobility aspects is taken into account via the net internal and domestic migration per 1,000 inhabitants. This diverse set of variables ensures a robust framework for understanding the factors influencing economic dynamics at the county level and serves the scope for showcasing the data leakage consequences of an erroneous split of the training-test data.

---

[23] [23] Personal income includes the total income received by all individuals and entities in a county from all sources (sum of wages and salaries, supplements to wages and salaries, proprietors' income with inventory valuation and capital consumption adjustments, rental income of persons with capital consumption adjustment, personal dividend income, personal interest income, and personal current transfer receipts, less contributions for government social insurance plus the adjustment for residence), which is then divided by the number of individuals (both civilian and military) who reside in the county.

**Table A.1: Variable details**

| Variable | Source |
| --- | --- |
| Personal income per capita growth rate (%) | U.S. Bureau of Economic Analysis (BEA) |
| Recession dummy | U.S. Bureau of Economic Analysis (BEA) |
| Personal income per capita | U.S. Bureau of Economic Analysis (BEA) |
| Average wage | U.S. Bureau of Economic Analysis (BEA) |
| Income from Unemployment Benefit (%) | U.S. Bureau of Economic Analysis (BEA) |
| Workplace employment rate (%) | U.S. Bureau of Economic Analysis (BEA) |
| Population | U.S. Census data |
| Population under 18 (%) | U.S. Census data |
| Population over 65 (%) | U.S. Census data |
| Women (% of population) | U.S. Census data |
| White (% of population) | U.S. Census data |
| Black (% of population) | U.S. Census data |
| Hispanic (% of population) | U.S. Census data |
| Asian (% of population) | U.S. Census data |
| Birth (per 1,000 inhabitants) | U.S. Census data |
| Deaths (per 1,000 inhabitants) | U.S. Census data |
| Net internal migration (per 1,000 inhabitants) | U.S. Census data |
| Net domestic migration (per 1,000 inhabitants) | U.S. Census data |

**Table A.2: Descriptive statistics**

| Variable | Mean | Std Dev |
|---|---|---|
| Personal income per capita growth rate (%) | 3.38 | 5.66 |
| Recession dummy | 0.15 | 0.36 |
| Personal income per capita | 34,089 | 11,280 |
| Average wage | 34,140 | 9,236 |
| Income from Unemployment Benefit (%) | 46.43 | 40.02 |
| Workplace employment rate (%) | 86.65 | 27.59 |
| Population | 98,965 | 317,471 |
| Population under 18 (%) | 23.53 | 3.40 |
| Population over 65 (%) | 16.51 | 4.54 |
| Women (% of population) | 50.13 | 2.07 |
| White (% of population) | 78.61 | 19.37 |
| Black (% of population) | 8.75 | 14.35 |
| Hispanic (% of population) | 8.12 | 13.13 |
| Asian (% of population) | 3.35 | 4.89 |
| Birth (per 1,000 inhabitants) | 11.16 | 3.80 |
| Deaths (per 1,000 inhabitants) | 9.57 | 3.57 |
| Net internal migration (per 1,000 inhabitants) | 1.00 | 1.82 |
| Net domestic migration (per 1,000 inhabitants) | -0.99 | 18.54 |
| | | |
| N | 3,058 | |
| T | 20 | |
| N·T | 61,160 | |

**Table A.3: Temporal leakage in the forecasting classification problem (all years) – Full Results**

| Model | Contemporaneous | Outcome Lags | Split Type | AUC | Adj. Test Size | Sensitivity | Specificity | Train Size | Test Size |
|---|---|---|---|---|---|---|---|---|---|
| logit | yes | yes | time | 0.335 | yes | 0.000 | 0.998 | 34244 | 2448 |
| logit | yes | yes | time | 0.338 | no | 0.000 | 1.000 | 42812 | 12232 |
| logit | no | yes | time | 0.357 | yes | 0.009 | 0.991 | 34244 | 2448 |
| logit | no | yes | time | 0.361 | no | 0.000 | 1.000 | 42812 | 12232 |
| logit | yes | no | time | 0.361 | yes | 0.997 | 0.001 | 34244 | 2448 |
| logit | yes | no | time | 0.377 | no | 0.999 | 0.000 | 42812 | 12232 |
| logit | no | no | time | 0.392 | yes | 0.997 | 0.003 | 34244 | 2448 |
| logit | yes | yes | random | 0.399 | no | 0.000 | 1.000 | 42934 | 12110 |
| logit | no | no | time | 0.409 | no | 0.000 | 1.000 | 42812 | 12232 |
| logit | yes | yes | random | 0.414 | yes | 0.079 | 0.937 | 34244 | 2448 |
| logit | no | yes | random | 0.423 | no | 0.096 | 0.911 | 42934 | 12110 |
| logit | no | yes | random | 0.437 | yes | 0.171 | 0.877 | 34244 | 2448 |
| logit | yes | no | random | 0.473 | no | 0.000 | 0.999 | 42934 | 12110 |
| logit | yes | no | random | 0.487 | yes | 0.661 | 0.355 | 34244 | 2448 |
| logit | no | no | random | 0.493 | no | 0.132 | 0.878 | 42934 | 12110 |
| logit | no | no | random | 0.508 | yes | 0.685 | 0.368 | 34244 | 2448 |
| xgboost | no | no | random | 0.694 | no | 0.661 | 0.622 | 42934 | 12110 |
| xgboost | no | yes | time | 0.695 | yes | 0.674 | 0.665 | 34244 | 2448 |
| rforest | no | no | time | 0.698 | yes | 0.680 | 0.650 | 34244 | 2448 |
| xgboost | no | no | time | 0.698 | yes | 0.759 | 0.571 | 34244 | 2448 |
| rforest | no | yes | time | 0.699 | yes | 0.710 | 0.619 | 34244 | 2448 |
| xgboost | yes | yes | time | 0.705 | yes | 0.564 | 0.793 | 34244 | 2448 |
| rforest | no | no | random | 0.705 | no | 0.646 | 0.647 | 42934 | 12110 |
| rforest | no | no | random | 0.711 | yes | 0.605 | 0.692 | 34244 | 2448 |
| xgboost | no | no | random | 0.712 | yes | 0.568 | 0.743 | 34244 | 2448 |
| xgboost | yes | no | time | 0.713 | yes | 0.649 | 0.722 | 34244 | 2448 |
| xgboost | no | yes | time | 0.713 | no | 0.649 | 0.701 | 42812 | 12232 |
| xgboost | no | yes | random | 0.714 | no | 0.587 | 0.724 | 42934 | 12110 |
| xgboost | no | no | time | 0.715 | no | 0.632 | 0.714 | 42812 | 12232 |
| rforest | no | no | time | 0.717 | no | 0.745 | 0.593 | 42812 | 12232 |
| rforest | no | yes | time | 0.718 | no | 0.733 | 0.608 | 42812 | 12232 |
| xgboost | yes | yes | time | 0.719 | no | 0.611 | 0.744 | 42812 | 12232 |
| xgboost | yes | no | time | 0.722 | no | 0.650 | 0.706 | 42812 | 12232 |
| rforest | yes | no | time | 0.723 | yes | 0.689 | 0.681 | 34244 | 2448 |
| xgboost | no | yes | random | 0.724 | yes | 0.598 | 0.736 | 34244 | 2448 |
| rforest | yes | yes | time | 0.734 | yes | 0.723 | 0.648 | 34244 | 2448 |
| rforest | yes | no | time | 0.736 | no | 0.724 | 0.644 | 42812 | 12232 |
| rforest | yes | yes | time | 0.745 | no | 0.625 | 0.748 | 42812 | 12232 |
| rforest | no | yes | random | 0.749 | no | 0.624 | 0.743 | 42934 | 12110 |
| rforest | no | yes | random | 0.757 | yes | 0.661 | 0.722 | 34244 | 2448 |
| xgboost | yes | no | random | 0.769 | no | 0.662 | 0.743 | 42934 | 12110 |
| xgboost | yes | yes | random | 0.787 | no | 0.756 | 0.671 | 42934 | 12110 |
| xgboost | yes | no | random | 0.789 | yes | 0.668 | 0.780 | 34244 | 2448 |
| rforest | yes | no | random | 0.796 | no | 0.740 | 0.714 | 42934 | 12110 |
| rforest | yes | no | random | 0.803 | yes | 0.764 | 0.711 | 34244 | 2448 |
| xgboost | yes | yes | random | 0.804 | yes | 0.629 | 0.846 | 34244 | 2448 |
| rforest | yes | yes | random | 0.820 | no | 0.720 | 0.761 | 42934 | 12110 |
| rforest | yes | yes | random | 0.830 | yes | 0.757 | 0.757 | 34244 | 2448 |

*Notes:* the table reports the details of the results shown in Figure 2 – Panel A.

**Table A.4: Temporal leakage in the forecasting regression problem (all years) – Full Results**

| Model | Contemporaneous | Outcome Lags | Split Type | MSE | Adj. Test Size | Train Size |
|---|---|---|---|---|---|---|
| 2448 | yes | yes | 0.0421066 | yes | 34244 | yes |
| 12232 | yes | yes | 0.0436146 | no | 42812 | yes |
| 2448 | no | yes | 0.044298 | yes | 34244 | no |
| 12232 | yes | yes | 0.0445243 | no | 42812 | yes |
| 2448 | yes | yes | 0.0454925 | yes | 34244 | yes |
| 12232 | no | yes | 0.045622 | no | 42812 | no |
| 12232 | no | yes | 0.0469008 | no | 42812 | no |
| 2448 | no | yes | 0.0481031 | yes | 34244 | no |
| 2448 | yes | yes | 0.0481564 | yes | 34244 | yes |
| 2448 | yes | yes | 0.0488085 | yes | 34244 | yes |
| 2448 | yes | yes | 0.0491333 | yes | 34244 | yes |
| 12110 | yes | yes | 0.0494632 | no | 42934 | yes |
| 12110 | yes | yes | 0.0504076 | no | 42934 | yes |
| 12232 | yes | yes | 0.0506797 | no | 42812 | yes |
| 2448 | yes | yes | 0.0510477 | yes | 34244 | yes |
| 12110 | yes | yes | 0.0515274 | no | 42934 | yes |
| 2448 | no | yes | 0.0521848 | yes | 34244 | yes |
| 2448 | no | yes | 0.0530976 | yes | 34244 | yes |
| 12110 | no | yes | 0.0544512 | no | 42934 | yes |
| 12110 | no | yes | 0.0552178 | no | 42934 | yes |
| 2448 | no | yes | 0.0558515 | yes | 34244 | yes |
| 12110 | no | yes | 0.0569697 | no | 42934 | yes |
| 12232 | no | yes | 0.0603354 | no | 42812 | no |
| 2448 | no | yes | 0.0618772 | yes | 34244 | no |
| 2448 | yes | no | 0.0861945 | yes | 34244 | yes |
| 12110 | yes | no | 0.0876502 | no | 42934 | yes |
| 12110 | no | no | 0.0914819 | no | 42934 | yes |
| 2448 | no | no | 0.0923943 | yes | 34244 | yes |
| 2448 | yes | no | 0.1070117 | yes | 34244 | yes |
| 12110 | yes | no | 0.1072542 | no | 42934 | yes |
| 12110 | no | no | 0.1109047 | no | 42934 | yes |
| 2448 | no | no | 0.1123978 | yes | 34244 | yes |
| 2448 | no | no | 0.1234586 | yes | 34244 | no |
| 2448 | yes | no | 0.1243655 | yes | 34244 | yes |
| 12232 | yes | no | 0.1247234 | no | 42812 | yes |
| 12232 | no | no | 0.1248404 | no | 42812 | no |
| 12232 | yes | no | 0.1278797 | no | 42812 | yes |
| 2448 | yes | no | 0.1282474 | yes | 34244 | yes |
| 2448 | no | no | 0.1295913 | yes | 34244 | no |
| 12232 | no | no | 0.1303021 | no | 42812 | no |
| 12110 | yes | no | 0.1528064 | no | 42934 | yes |
| 2448 | yes | no | 0.1539675 | yes | 34244 | yes |
| 2448 | yes | no | 0.160084 | yes | 34244 | yes |
| 12232 | yes | no | 0.1616486 | no | 42812 | yes |
| 12110 | no | no | 0.1665273 | no | 42934 | yes |
| 2448 | no | no | 0.1682672 | yes | 34244 | no |
| 2448 | no | no | 0.1687337 | yes | 34244 | yes |
| 12232 | no | no | 0.1717883 | no | 42812 | no |

*Notes:* The table reports the details of the results shown in Figure 2 – Panel B.

**Table A.5: Temporal leakage in the forecasting classification problem (focus on the Great Recession, year 2009) – Full Results**

| Model | Contemporaneous | Outcome Lags | Split Type | AUC | Adj. Test Size | Sensitivity | Specificity | Train Size | Test Size |
|---|---|---|---|---|---|---|---|---|---|
| xgboost | no | no | time | 0.393 | yes | 0.020 | 1.000 | 14676 | 612 |
| xgboost | yes | no | time | 0.399 | yes | 0.046 | 0.968 | 14676 | 612 |
| rforest | no | no | time | 0.409 | yes | 0.033 | 0.991 | 14676 | 612 |
| xgboost | no | yes | time | 0.422 | yes | 0.025 | 0.995 | 14676 | 612 |
| xgboost | no | no | time | 0.424 | no | 0.016 | 0.991 | 18348 | 3058 |
| xgboost | yes | yes | time | 0.429 | yes | 0.079 | 0.950 | 14676 | 612 |
| xgboost | yes | no | time | 0.431 | no | 0.069 | 0.939 | 18348 | 3058 |
| rforest | no | yes | time | 0.440 | yes | 0.025 | 1.000 | 14676 | 612 |
| rforest | no | no | time | 0.441 | no | 0.021 | 0.985 | 18348 | 3058 |
| logit | yes | yes | random | 0.449 | yes | 0.082 | 0.936 | 14676 | 612 |
| rforest | yes | no | time | 0.462 | yes | 0.015 | 0.995 | 14676 | 612 |
| rforest | yes | yes | time | 0.463 | yes | 0.086 | 0.950 | 14676 | 612 |
| logit | yes | yes | random | 0.465 | no | 0.159 | 0.865 | 16696 | 4710 |
| xgboost | no | yes | time | 0.466 | no | 0.135 | 0.905 | 18348 | 3058 |
| logit | no | yes | random | 0.470 | yes | 0.082 | 0.944 | 14676 | 612 |
| xgboost | yes | yes | time | 0.472 | no | 0.074 | 0.962 | 18348 | 3058 |
| rforest | no | yes | time | 0.479 | no | 0.149 | 0.892 | 18348 | 3058 |
| rforest | yes | no | time | 0.485 | no | 0.267 | 0.746 | 18348 | 3058 |
| logit | no | yes | random | 0.488 | no | 0.263 | 0.787 | 16696 | 4710 |
| rforest | yes | yes | time | 0.496 | no | 0.157 | 0.893 | 18348 | 3058 |
| logit | yes | no | random | 0.554 | yes | 0.545 | 0.612 | 14676 | 612 |
| logit | yes | no | random | 0.564 | no | 0.481 | 0.649 | 16696 | 4710 |
| logit | no | no | random | 0.572 | yes | 0.545 | 0.627 | 14676 | 612 |
| logit | no | no | random | 0.582 | no | 0.537 | 0.622 | 16696 | 4710 |
| logit | yes | yes | time | 0.584 | no | 0.548 | 0.597 | 18348 | 3058 |
| logit | yes | yes | time | 0.588 | yes | 0.277 | 0.867 | 14676 | 612 |
| logit | no | yes | time | 0.588 | no | 0.556 | 0.594 | 18348 | 3058 |
| logit | no | yes | time | 0.592 | yes | 0.279 | 0.867 | 14676 | 612 |
| logit | yes | no | time | 0.638 | no | 0.640 | 0.566 | 18348 | 3058 |
| logit | no | no | time | 0.645 | no | 0.625 | 0.593 | 18348 | 3058 |
| logit | yes | no | time | 0.656 | yes | 0.751 | 0.482 | 14676 | 612 |
| logit | no | no | time | 0.664 | yes | 0.782 | 0.459 | 14676 | 612 |
| xgboost | no | no | random | 0.696 | no | 0.729 | 0.561 | 16696 | 4710 |
| rforest | no | no | random | 0.713 | yes | 0.645 | 0.663 | 14676 | 612 |
| rforest | no | no | random | 0.714 | no | 0.639 | 0.687 | 16696 | 4710 |
| xgboost | no | no | random | 0.718 | yes | 0.627 | 0.707 | 14676 | 612 |
| xgboost | no | yes | random | 0.745 | no | 0.674 | 0.714 | 16696 | 4710 |
| xgboost | no | yes | random | 0.758 | yes | 0.864 | 0.532 | 14676 | 612 |
| rforest | no | yes | random | 0.780 | no | 0.738 | 0.704 | 16696 | 4710 |
| rforest | no | yes | random | 0.792 | yes | 0.618 | 0.837 | 14676 | 612 |
| rforest | yes | no | random | 0.830 | yes | 0.727 | 0.787 | 14676 | 612 |
| rforest | yes | no | random | 0.840 | no | 0.747 | 0.812 | 16696 | 4710 |
| xgboost | yes | no | random | 0.844 | no | 0.752 | 0.814 | 16696 | 4710 |
| xgboost | yes | yes | random | 0.853 | yes | 0.682 | 0.884 | 14676 | 612 |
| xgboost | yes | yes | random | 0.854 | no | 0.711 | 0.865 | 16696 | 4710 |
| rforest | yes | yes | random | 0.861 | no | 0.764 | 0.830 | 16696 | 4710 |
| rforest | yes | yes | random | 0.862 | yes | 0.709 | 0.884 | 14676 | 612 |
| xgboost | yes | no | random | 0.865 | yes | 0.845 | 0.763 | 14676 | 612 |

*Notes:* The table reports the details of the results shown in Figure 3 – Panel A.

**Table A.6: Temporal leakage in the forecasting regression problem (focus on the Great Recession, year 2009) – Full Results**

| Model | Contemporaneous | Outcome Lags | Split Type | MSE | Adj. Test Size | Train Size | Test Size |
|---|---|---|---|---|---|---|---|
| rforest | yes | yes | random | 0.052 | no | 16696 | 4710 |
| ols | yes | yes | random | 0.053 | no | 16696 | 4710 |
| xgboost | yes | yes | random | 0.053 | no | 16696 | 4710 |
| rforest | yes | yes | random | 0.054 | yes | 14676 | 612 |
| xgboost | yes | yes | random | 0.055 | yes | 14676 | 612 |
| rforest | no | yes | random | 0.056 | no | 16696 | 4710 |
| ols | yes | yes | random | 0.058 | yes | 14676 | 612 |
| ols | no | yes | random | 0.058 | no | 16696 | 4710 |
| rforest | no | yes | random | 0.058 | yes | 14676 | 612 |
| ols | yes | yes | time | 0.060 | no | 18348 | 3058 |
| xgboost | no | yes | random | 0.061 | no | 16696 | 4710 |
| ols | no | yes | random | 0.063 | yes | 14676 | 612 |
| ols | yes | yes | time | 0.065 | yes | 14676 | 612 |
| xgboost | no | yes | random | 0.066 | yes | 14676 | 612 |
| xgboost | yes | yes | time | 0.074 | yes | 14676 | 612 |
| xgboost | yes | yes | time | 0.076 | no | 18348 | 3058 |
| rforest | yes | yes | time | 0.086 | yes | 14676 | 612 |
| rforest | yes | yes | time | 0.086 | no | 18348 | 3058 |
| ols | no | yes | time | 0.086 | no | 18348 | 3058 |
| xgboost | yes | no | random | 0.086 | no | 16696 | 4710 |
| ols | no | yes | time | 0.088 | yes | 14676 | 612 |
| xgboost | no | no | random | 0.089 | no | 16696 | 4710 |
| xgboost | yes | no | random | 0.090 | yes | 14676 | 612 |
| rforest | no | yes | time | 0.091 | yes | 14676 | 612 |
| rforest | no | yes | time | 0.092 | no | 18348 | 3058 |
| xgboost | no | no | random | 0.093 | yes | 14676 | 612 |
| xgboost | no | no | time | 0.095 | no | 18348 | 3058 |
| xgboost | no | yes | time | 0.097 | yes | 14676 | 612 |
| xgboost | no | yes | time | 0.097 | no | 18348 | 3058 |
| xgboost | no | no | time | 0.098 | yes | 14676 | 612 |
| xgboost | yes | no | time | 0.099 | no | 18348 | 3058 |
| rforest | yes | no | random | 0.104 | no | 16696 | 4710 |
| rforest | no | no | random | 0.104 | no | 16696 | 4710 |
| xgboost | yes | no | time | 0.105 | yes | 14676 | 612 |
| rforest | no | no | random | 0.108 | yes | 14676 | 612 |
| rforest | yes | no | random | 0.109 | yes | 14676 | 612 |
| rforest | no | no | time | 0.111 | no | 18348 | 3058 |
| rforest | yes | no | time | 0.114 | yes | 14676 | 612 |
| rforest | yes | no | time | 0.114 | no | 18348 | 3058 |
| rforest | no | no | time | 0.114 | yes | 14676 | 612 |
| ols | yes | no | random | 0.143 | no | 16696 | 4710 |
| ols | no | no | time | 0.143 | no | 18348 | 3058 |
| ols | no | no | time | 0.144 | yes | 14676 | 612 |
| ols | no | no | random | 0.148 | no | 16696 | 4710 |
| ols | yes | no | random | 0.150 | yes | 14676 | 612 |
| ols | yes | no | time | 0.152 | no | 18348 | 3058 |
| ols | no | no | random | 0.156 | yes | 14676 | 612 |
| ols | yes | no | time | 0.156 | yes | 14676 | 612 |

*Notes:* The table reports the details of the results shown in Figure 3 – Panel B.

**Table A.7: Cross-Sectional leakage in the classification problem (all years) – Full Results**

| Model | Split Type | AUC | Adj. Test Size | Sensitivity | Specificity | Train Size | Test Size |
|---|---|---|---|---|---|---|---|
| logit | space | 0.404 | yes | 0.000 | 0.999 | 34244 | 2448 |
| logit | space | 0.422 | no | 0.060 | 0.945 | 44028 | 11016 |
| logit | random | 0.423 | no | 0.096 | 0.911 | 42934 | 12110 |
| logit | random | 0.437 | yes | 0.171 | 0.877 | 34244 | 2448 |
| logit | space_states | 0.456 | yes | 0.112 | 0.900 | 34244 | 2448 |
| logit | space_states | 0.466 | no | 0.128 | 0.899 | 44208 | 10836 |
| xgboost | space | 0.686 | yes | 0.560 | 0.705 | 34244 | 2448 |
| xgboost | space_states | 0.693 | yes | 0.529 | 0.749 | 34244 | 2448 |
| xgboost | space_states | 0.697 | no | 0.585 | 0.695 | 44208 | 10836 |
| xgboost | space | 0.705 | no | 0.710 | 0.580 | 44028 | 11016 |
| rforest | space_states | 0.708 | no | 0.588 | 0.717 | 44208 | 10836 |
| rforest | space_states | 0.713 | yes | 0.620 | 0.704 | 34244 | 2448 |
| xgboost | random | 0.714 | no | 0.587 | 0.724 | 42934 | 12110 |
| xgboost | random | 0.724 | yes | 0.598 | 0.736 | 34244 | 2448 |
| rforest | space | 0.734 | yes | 0.721 | 0.631 | 34244 | 2448 |
| rforest | random | 0.749 | no | 0.624 | 0.743 | 42934 | 12110 |
| rforest | random | 0.757 | yes | 0.661 | 0.722 | 34244 | 2448 |
| rforest | space | 0.758 | no | 0.687 | 0.697 | 44028 | 11016 |

*Notes:* The table reports the details of the results shown in Figure 4 – Panel A.

**Table A.8: Cross-Sectional leakage in the regression problem (all years) – Full Results**

| Model | Contemporaneous | Outcome Lags | Split Type | MSE | Adj. Test Size | Train Size | Test Size |
|---|---|---|---|---|---|---|---|
| random | no | yes | random | 0.052 | yes | 34244 | 2448 |
| space | no | yes | space | 0.052 | yes | 34244 | 2448 |
| random | no | yes | random | 0.053 | yes | 34244 | 2448 |
| space | no | yes | space | 0.053 | no | 44028 | 11016 |
| space | no | yes | space | 0.054 | yes | 34244 | 2448 |
| random | no | yes | random | 0.054 | no | 42934 | 12110 |
| space | no | yes | space | 0.055 | no | 44028 | 11016 |
| random | no | yes | random | 0.055 | no | 42934 | 12110 |
| random | no | yes | random | 0.056 | yes | 34244 | 2448 |
| space | no | yes | space | 0.056 | yes | 34244 | 2448 |
| random | no | yes | random | 0.057 | no | 42934 | 12110 |
| space | no | yes | space | 0.058 | no | 44028 | 11016 |
| space_states | no | yes | space_states | 0.062 | no | 44208 | 10836 |
| space_states | no | yes | space_states | 0.063 | no | 44208 | 10836 |
| space_states | no | yes | space_states | 0.063 | yes | 34244 | 2448 |
| space_states | no | yes | space_states | 0.065 | yes | 34244 | 2448 |
| space_states | no | yes | space_states | 0.072 | yes | 34244 | 2448 |
| space_states | no | yes | space_states | 0.072 | no | 44208 | 10836 |

*Notes:* The table reports the details of the results shown in Figure 4 – Panel B.

B. **Machine Learning pipeline**

- **B.1. Machine Learning models**

Below we report a brief description of the models used in the empirical application. Please refer to Hastie et al. (2009) for additional details regarding the algorithms

- **Random Forest** is an ensemble learning technique that constructs many decision trees during training and aggregates their predictions to increase out-of-sample accuracy by reducing overfitting risk. Each tree is built using a random subset of the training data, which helps in reducing variance and making the model more robust. Additionally, a second layer of randomness is introduced by forcing the trees to select among and split on only a random subset of the predictors at each candidate split. The final prediction is determined by averaging the outputs of all the trees for regression tasks and by taking a majority vote for classification tasks. We use the following parameters: trees =500, maximum depth of each tree = none; minimum samples per leaf =4; maximum number of candidate variables at each split = square root of the number of predictors; and minimum number of samples per split = 10.

- **Extreme Gradient Boosting (XGBoost)** is a tree-based ensemble technique that builds models in a sequential manner, where each new model tries to predict the residuals, i.e., the errors, of its predecessors. XGBoost improves this approach by optimizing computational efficiency through parallelization, using regularization to prevent overfitting, and handling missing data. It is widely used in ML applications on structured/tabular data due to its ability to handle complex datasets with minimal tuning. We use it with the following parameters: learning rate= 0.01; maximum depth of each tree = 2; sum of instance weight (hessian) needed in a child = 5; minimum loss reduction required to make a further partition on a leaf node of the tree (gamma) = 1; subsample size = 0.5; subsample ratio of columns when constructing each tree = 0.8; and 500 boost rounds.

- **B. 2. Variables used in the models**

Across all models, we use the following variables as predictors:

- **Group a)**: *Average wage, Income from Unemployment Benefit (%), Workplace employment rate (%), Population, Population under 18 (%), Population over 65 (%), Women (% of population), White (% of population), Black (% of population), Hispanic (% of population), Asian (% of population), Birth (per 1,000 inhabitants), Deaths (per 1,000 inhabitants), Net* *internal migration (per 1,000 inhabitants) and Net domestic migration (per 1,000 inhabitants)*

- **Group b)** *(%), Recession dummy, Log of Personal income per capita*, (%) GDP growth,

Note that when running a model that includes contemporaneous predictors, we use all variables in *Group a)* but not those of *Group b),* since these contains the target variable or a direct transformation of it. In all models, all the variables of *Group a)* and *Group b)* described enter the model with *lag t-1 and t-2*, except for the outcome variable, which enters the model only when we include the lagged outcome.